\input harvmac
\input tables

\font\zfont = cmss10 
\def\ZZ{\hbox{\zfont Z\kern-.4emZ}}

\def\nc{{N_{c}}}

\def\nfl{{N_{f}}}\def\nfr{{N'_{f}}}
\def\ncl{{\nc}}\def\ncr{{N'_{c}}}
\def\ncld{{\tilde{N}_c}}\def\ncrd{{\tilde{N}'_c}}

%

\def\Ql{Q}\def\Qr{Q'}

\def\Qlt{\tilde{\Ql}}\def\Qrt{\tilde{\Qr}}

%
%

\def\ql{q}\def\qr{\ql'}
\def\qlt{\tilde\ql}\def\qrt{\tilde\qr}

\noblackbox
\def\np#1#2#3{Nucl. Phys. {\bf B#1} (#2) #3}
\def\pl#1#2#3{Phys. Lett. {\bf #1B} (#2) #3}

\def\physrev#1#2#3{Phys. Rev. {\bf D#1} (#2) #3}

\def\prep#1#2#3{Phys. Rep. {\bf #1} (#2) #3}

\def\Tr{{\rm Tr ~}}

\def\tilde{\widetilde}

\Title{hep-th/9605232, PU-1626}
{\vbox{\centerline{Duality in Supersymmetric SU($N_c$) Gauge Theory}
 \centerline{with Two Adjoint Chiral Superfields}}}
\bigskip
\centerline{\it John Brodie}
\smallskip
\centerline{Department of Physics}
\centerline{Princeton University}
\centerline{Princeton, NJ 08540, USA}
\centerline{jhbrodie@princeton.edu}
\vskip .2in

\vglue .3cm

\bigskip\bigskip

\noindent
We discuss $SU(N_c)$ gauge theory coupled to two adjoint chiral
superfields $X$ and $Y$, and a number of fundamental chiral
superfields $Q^i$. We add a superpotential that has the form of
Arnold's $D$ series $W = \Tr X^{k+1} + \Tr XY^2$.  We present a dual
description in terms of an $SU(3kN_f - N_c)$ gauge theory, and we show
that the duality passes many tests.  At the end of the paper, we show
how a deformation of this superpotential flows to another duality
having a product gauge group $SU(N_c)\times SU(N_c')$, with an adjoint
field charged under $SU(N_c)$, an adjoint field charged under
$SU(N_c')$, fields in the $(N_c,N_c')$ and $(\overline N_c,\overline
N_c')$ representation, and a number of fundamentals. The dual
description is an $SU(2kN_f' + kN_f - N_c')\times SU(2kN_f + kN_f' -
N_c)$ gauge theory.
\Date{6/96}

\nref\ads{I. Affleck, M. Dine, and N. Seiberg, \np{241}{1984}{493};
\np{256}{1985}{557}}%
\nref\nsvz{V.A. Novikov, M.A. Shifman, A. I.  Vainstein and V. I.
Zakharov, \np{223}{1983}{445}; \np{260}{1985}{157};
 \np{229}{1983}{381}}
\nref\svholo{M.A. Shifman and A.I Vainshtein, \np{277}{1986}{456};
\np{359}{1991}{571}}%
\nref\cern{D. Amati, K. Konishi, Y. Meurice, G.C. Rossi and G.
Veneziano, \prep{162}{1988}{169} and references therein}%
\nref\nonren{N. Seiberg,                 \pl{318}{1993}{469}}%
\nref\natii{N. Seiberg,                 \physrev{49}{1994}{6857}}%
\nref\ils{K. Intriligator, R. Leigh and N. Seiberg,
\physrev{50}{1994}{1092}
}%
\nref\sw{N. Seiberg and E. Witten, \np{426}{1994}{19};
                \np{431} {1994} {484}
}%
\nref\intse{K. Intriligator and N. Seiberg,
\np{431} {1994} {551}
}%

\nref\nati{N. Seiberg, \np {435}{1995}{129}}%

\nref\isson{K. Intriligator and N. Seiberg, {\it Duality,
Monopoles,  Dyons, Confinement and Oblique Confinement in
Supersymmetric $SO(N_c)$ Gauge Theories}, RU--95--3,
hep-th/9503179, \np{444} {1995} {125}.}

\nref\intpou{K. Intriligator and P. Pouliot, {\it Exact
Superpotentials, Quantum Vacua and Duality in Supersymmetric
$Sp(N_c)$ Gauge Theories}, RU--95--23, hep-th/9505006, \np{353} 
{1995} {471}.}

\nref\poul{P. Pouliot, hep-th/9507018, RU-95-46, \np{359} {1995} 108.}%

\nref\poulstr{P. Pouliot and M.J. Strassler, hep-th/9510228, RU-95-67,
\np{370} {1996} 76.}%

\nref\poulstra{P. Pouliot and M.J. Strassler, hep-th/9507018, RU-95-78}%

\nref\dk{D. Kutasov, EFI-95-11, hep-th/9503086,\pl{351}{1995} 230.}%

\nref\kut{D. Kutasov and A. Schwimmer, hep-th/9505004, \pl{354} {1995} 315.}

\nref\kss{D. Kutasov, A. Schwimmer, and N. Seiberg, hep-th/9510222, 
\np{459}{1996} 455}.

\nref\berk{M. Berkooz, {\it The Dual of Supersymmetric SU(2k)
with an Antisymmetric Tensor and Composite Dualities}, RU-95-20,
hep-th/9505088.}

\nref\emop{R.G. Leigh and M.J. Strassler,
{\it Exactly Marginal Operators and Duality in
Four Dimensional N=1 Supersymmetric Gauge Theory},
RU--95--2, hep-th/9503121, \np{447} {1995} 95.}

\nref\rlmsspso{R.G. Leigh and M.J. Strassler, {\it Duality of
$Sp(2N_c)$ and $SO(N_c)$ Supersymmetric Gauge Theories with
Adjoint Matter}, RU--95--30, hep-th/9505088, \pl {356} {1995} 492.}

\nref\kispso{K. Intriligator, {\it New RG Fixed Points and
Duality in Supersymmetric $Sp(N_c)$ and $SO(N_c)$ Gauge
Theories}, RU--95--27, hep-th/9505051, to appear in Nucl. Phys.
B.}

\nref\Ahsonyank{O. Aharony, J. Sonnenschein, and S. Yankielowicz,
{\it Flows and Duality Symmetries in N=1 Supersymmetric Gauge
Theories}, TAUP--2246--95, CERN-TH/95--91, hep-th/9504113, 
\np{449} {1995} 509.}

\nref\ilst{K. Intriligator, R. Leigh and M. Strassler, hep-th/9506148,
RU-95-38, \np{456}{1995}567.}%

\newsec{Introduction.}

Drawing on the work of \refs{\ads - \intse}, N. Seiberg was able to
show that two different four dimensional $N=1$ supersymmetric gauge
theories can flow to the same non-trivial fixed point or that a theory
can flow to a dual free magnetic phase \nati. Since then, many more
examples of such dualities have been discovered with matter in the
fundamental representation of the gauge group \refs{\isson - \intpou}.
A few duals have been found for chiral theories involving more matter
than just fundamentals \refs{\poul - \poulstra}.  In general, theories
with two index tensors have proved difficult to solve.  For example, a
general dual theory was searched for in the case of $SU(N_c)$ gauge
theory with $N_f$ fundamentals and one adjoint, $X$, but to no avail.
Instead, a special version of the theory was solved by Kutasov et
al. \refs{\dk - \kss}\ by adding a superpotential of the form
\eqn\wk{W=\Tr X^{k+1}.}
This superpotential does three things. It reduces the number of gauge
invariant operators in the theory by setting $X^k = 0$, it fixes the
R-charge of the field $X$, and, because $X^{k+1}$ is a dangerously
irrelevant operator, it presumably takes the theory from the poorly 
understood fixed
point, to a new fixed point that can be solved. Since the
solution of this one adjoint model, other examples of dualities involving one
two-index tensor were found \refs{\berk-\ilst}.  In this paper, like
Kutasov et al., we will employ a superpotential, but here we have
fundamentals and {\it two} adjoint fields. The superpotential we will
add looks like
\eqn\b{W=\Tr X^{k+1} + \Tr XY^2.}
The dual gauge group is $SU(3kN_f-N_c)$.  It is interesting to note
that the form of the superpotential looks like the $D_{k+2}$
classification of Arnold's singularity theory
\nref\arn{V.I. Arnold,{\it Singularity Theory} London Mathematical Lecture
Notes Series: 53, Cambridge University Press (1981)}.  However, in our
case the $X$ and $Y$ field carry gauge indices and therefore do not
commute.  Nevertheless, it is tempting to speculate that perhaps there
is a relation between singularity theory and conformal field theories
in four dimensions as there is in two dimensions. Such a connection
between Arnold's $A_k$ classification and the superpotential \wk\ was
exploited in \kss.

At the end of this paper, we will show how the superpotential can be
deformed and flow to a prediction for another duality with a product
gauge group $SU(N_c)\times SU(N_c')$ and a superpotential
\eqn\supF{\eqalign{W=\Tr X_1^{k+1} + \Tr X_2^{k+1}
+ \Tr X_1F\tilde F + \Tr X_2F\tilde F.\cr}} where adjoint field $X_1$
is charged under $SU(N_c)$, adjoint field $X_2$ is charged under
$SU(N_c')$, and fields $F$ and $\tilde F$ are in the $(N_c,N_c')$
representation. There are also $N_f$ fundamentals and $N_f$
anti--fundamentals $Q$ and $\tilde Q$ that are charged under only
$SU(N_c)$ and $N_f$ fundamentals and $N_f$ anti--fundamentals $Q'$ and
$\tilde Q'$ that are charged under only $SU(N_c')$.  The gauge group
that we propose as dual to this theory is $SU(2kN_f' + kN_f -
N_c')\times SU(2kN_f + kN_f' - N_c)$

\newsec{The Theory}

The theory we will discuss in this paper will be a supersymmetric 
non-Abelian gauge theory with gauge group $SU(N_c)$, two chiral 
superfields $X$ and $Y$ which transform under the adjoint representation 
of the gauge group, $N_f$ fundamentals $Q^i$, $N_f$ anti--fundamentals 
$\tilde Q^i$ where $i=1, \ldots, N_f$. The theory is asymptotically free 
for $N_c>N_f$. We will also add a superpotential of the form 
\eqn\b{W={s_1\over k+1}\Tr X^{k+1} + s_2\Tr XY^2 + \lambda_1 \Tr X
+ \lambda_2 \Tr Y} where $\lambda_1$ and $\lambda_2$ are Lagrange
multipliers to implement the tracelessness condition on the adjoint
matter fields $X$ and $Y$.  This superpotential \b\ is not corrected
by non-perturbative effects for all values of $N_f$ \nonren .  The
equations of motion that follow from this superpotential are
\eqn\eom{\eqalign{s_1 X^k + s_2 Y^2 + \lambda_1 = 0\cr 
s_2(XY+YX) + \lambda_2 = 0. \cr}} These equations truncate the chiral
ring for odd and values of $k$. To illustrate this we can ignore
$\lambda_1$ and $\lambda_2$; statements below will then be modulo
lower order terms already included in the chiral ring.  We can
multiply $X^k$ by $Y$ from the right or left, and use the first
equation in \eom\ to show that
\eqn\ring{YX^k + X^kY = -2{s_2\over s_1}Y^3.}
Now, we can use the second equation in \eom\ to anticommute the $Y$
fields through the $X$ fields.
\eqn\ring{((-1)^k+ 1)X^kY = -2{s_2\over s_1}Y^3}
Thus for odd $k$, $Y^3=0$. The chiral ring is now said to be truncated
because we can relate higher order operators to lower ones.  {}From
$Y^3 = 0$ and equation \eom, it follows that in theories with
superpotentials having odd values of $k$ for $k<N_c$, all gauge
invariants can be formed from products of the $Q$, the $\tilde Q$, and
$X^{l-1} Y^{j-1}$ where $l=1,2,\ldots, k;\;j=1,2,3$. We can make
mesons of the form
\eqn\mlj{(M_{lj})_{\tilde i}^i=
\tilde Q_{\tilde i} X^{l-1} Y^{j-1} Q^i;\;\;\;l=1,2,\cdots, k;\;j=1,2,3}
We can also form baryons by introducing dressed quarks 
\eqn\j{Q_{(l,j)}=X^{l-1}Y^{j-1}Q;\;\;\;l=1,\cdots k;\;j=1,2,3.} 
and then contracting the gauge indices on an epsilon tensor.
\eqn\baryon{B^{(n_{1,1},n_{2,1},\cdots,n_{k,3})}=Q_{(1,1)}^{n_{1,1}}\cdots 
Q_{(k,3)}^{n_{k,3}} ;\;\;\sum_{l=1}^{k} \sum_{j=1}^{3} n_{l,j}=N_c}
The total number of baryons is
\eqn\totnum{\sum_{\{n_{l,j}\}}{N_f\choose n_{1,1}}\cdots{N_f\choose n_{k,3}}=
{3kN_f\choose N_c}.}  We can also make invariants of the form $\Tr
X^{l-1} Y^{j-1}$.  When $j=2$, or $j=3$ and $l$ is even, the trace
operators can be set to zero by using the equations of motion and the
cyclic property of traces.

\newsec{Duality}

The theory that we are proposing as dual to the one described above has an 
$SU(3kN_f-N_c)$ gauge group, 
two chiral 
superfields $ \bar X$ and $ \bar Y$ which transform under the 
adjoint representation 
of the gauge group, $N_f$ fundamentals $q^i$, $N_f$ 
anti--fundamentals 
$\tilde q^i$ where $i=1, \cdots, N_f$, and $M_{lj}$ singlets which are in a 
one-to-one mapping with the mesons of equation \mlj. 
The dual superpotential has the form 
\eqn\f{W=s_1\Tr \bar X^{k+1} + s_2\Tr \bar X \bar Y^2 +
{s_1 s_2\over \mu^4} \sum_{l=1}^{k} \sum_{j=1}^{3} M_{lj}\tilde q
 \bar X^{k-l} \bar Y^{3-j} q.}
Both theories have the following
 anomaly free global symmetries:
\eqn\globsym{SU(N_f)\times SU(N_f)\times U(1)_B\times U(1)_R}
The chiral superfields of the electric theory transform as follows under 
these global symmetries:
\eqn\d{\eqalign{
Q &\qquad (N_f,1,1, 1-{N_c\over N_f (k+1)}) \cr
\tilde Q & \qquad (1, \overline N_f,-1,1-{N_c\over N_f (k+1)})\cr
X &\qquad (1,1,0, {2\over k+1})\cr Y &\qquad (1,1,0, {k\over k+1}) .\cr
}}
The dual matter fields transform as follows under the global 
symmetries \globsym:
\eqn\e{\eqalign{
q &\qquad ( \overline N_f,1,{N_c\over 3kN_f-N_c}, 1-{3kN_f-N_c\over
 N_f (k+1)})
\cr
\tilde q & \qquad (1, N_f,-{N_c\over 3kN_f-N_c},1-{
3kN_f-N_c\over N_f (k+1)}) \cr
\overline X &\qquad (1,1,0, {2\over {k+1}}) \cr
\overline Y &\qquad (1,1,0, {k\over {k+1}}) \cr
M_{lj} &\qquad (N_f, \overline N_f,0, 2-{2\over {k+1}}{N_c\over
N_f}+{2(l-1)+k(j-1)\over {k+1}}) \cr }}
Again $l\le k$ and $j\le 3$.
This dual theory is asymptotically free 
for $3kN_f - N_c>N_f$ otherwise it is in a free magnetic phase. 
The superpotential \f\ is invariant under the flavor symmetries and baryon 
number. This is a non-trivial check of the mapping of the mesons 
\mlj\ into the singlet fields $M_{lj}$.
For the superpotential be invariant under the R-symmetry, it was essential 
to have 
the chiral ring truncated as described above and have the mapping of mesons
to elementary singlet fields as described in equation \mlj.
Baryon-like objects are mapped to other baryon-like objects in the dual theory.
The mapping is
\eqn\barmap{\eqalign{B_{\rm el}^{(n_{1,1},n_{2,1},\cdots,n_{k,3})}
\leftrightarrow
B_{\rm mag}^{(m_{1,1},m_{2,1},\cdots,m_{k,3})};\;\;
m_{l,j}=N_f-n_{k+1-l,4-j};\cr
l=1,2,\cdots, k;\;j=1,2,3\cr }}
The fact that this mapping is consistent with all global symmetries is another
non-trivial test of the proposed duality. 
Traces of products of $X$ and $Y$ are mapped to the same traces of products 
of $\bar X$ and $\bar Y$.

The t'Hooft anomaly matching conditions are satisfied for arbitrary $k$. 
They are 
\eqn\g{\eqalign{
SU(N_f)^3 \qquad &N_c d^{(3)}(N_f) \cr
SU(N_f)^2U(1)_R \qquad & -{1\over k+1}{N_c^2 \over N_f}d^{(2)}(N_f) \cr
SU(N_f)^2U(1)_B \qquad & N_cd^{(2)}(N_f) \cr
U(1)_R \qquad &-{1\over k+1}(N_c^2+1) \cr
U(1)_R^3 \qquad &\left(({2\over
k+1}-1)^3 +({k\over k+1}-1)^3 + 1\right)(N_c^2-1)-{2\over(k+1)^3}
{N_c^4\over N_f^2} \cr
U(1)_B^2U(1)_R \qquad &-{2\over k+1}N_c^2 . \cr}}
The t'Hooft anomaly matching conditions, especially the $U(1)_R^3$ condition,
are very powerful tests of the duality.

By considering only the global symmetries, we can find
a scale matching relation between the electric theory scale $\Lambda$ and the
magnetic theory scale $\bar\Lambda$ involving only the dimensionful parameter
$\mu$ and the coupling constants $s_1$ and $s_2$.
\eqn\scmat{\Lambda^{N_c-N_f}\bar \Lambda^{\bar N_c-N_f}
=Cs_1^{-3N_f}s_2^{-3kN_f}\mu^{4N_f}}
where $C$ is some constant. Such scale matching relations are common in
dual theories.

\subsec{k=1.}
Before we start deforming the superpotential, let's consider
the $k=1$ case. {}From the superpotential,
\eqn\supone{W={s_1\over 2}\Tr X^{2} + s_2\Tr XY^2 + \lambda_1 \Tr X
+ \lambda_2 \Tr Y,}
we see that $X$ is massive and below the scale $s_1$, we can integrate it out.
Using the equations of motion for $X$,
\eqn\eomX{s_1X + s_2Y^2 + \lambda_1,}
we see that the superpotential becomes
\eqn\supfour{W={3s_2^2\over 2s_1}\Tr Y^{4} + {2s_2\over s_1}\lambda_1\Tr Y^2
+ \lambda_2 \Tr Y.}
We recall that $\lambda_1$ is a Lagrange multiplier which should be integrated out.
By taking the trace of \eomX, we see that 
\eqn\lamb{\lambda_1 = s_2 \Tr Y^2.}
Inserting this into \supfour\ , we find
\eqn\supmarg{W={3s_2^2\over 2s_1}\Tr Y^4 + {2s_2^2\over s_1}(\Tr Y^2)^2
+ \lambda_2 \Tr Y,}
which is actually a marginal deformation of Kutasov's $k=3$ duality \wk.
Mesons involving the $X$ field are massive. Hence, we will be left
with only $Q\tilde Q$,$QY\tilde Q$,$QY^2\tilde Q$, and a dual 
$SU(3N_f-N_c)$ gauge group which is correct for
Kutasov's duality \kut.
The scale matching relations \scmat\ flows to the scale matching 
found in \kss
\eqn\kutscmat{\Lambda^{2N_c-N_f}\bar \Lambda^{\bar 2N_c-N_f}=
C\left( {\mu_L\over s_0}\right) ^{2N_f},}
where $s_0 = {s_2^2\over s_1}$ from \supmarg\ and
$\mu_L^2 = {s_0\mu^4\over s_1s_2}$ from \f\ and \kss .

\subsec{Theories with superpotentials having even k}
As we saw in section 2, the theories with a superpotential with $k$ odd have
a chiral ring that is truncated by the classical equations of motion. This is 
not the case for theories with a superpotential having even values of $k$. 
Nevertheless, as we have seen in this section, all of the checks on the 
duality work for $k$ even or odd. Thus, it is tempting to speculate that the 
duality also holds for theories with even values of $k$. The chiral
 ring for the $k$ even theories might be truncated by some quantum 
mechanical mechanism. We will see in the 
next section that if we make the assumption that the duality hold for both even
and odd values of $k$, we will naturally be led to another 
duality. 

\newsec{Deformations.}
\subsec{Giving a mass to M.}
We will now check that the duality and the scale matching
relation \scmat\ are consistent 
with the renormalization group flow
from a theory with $N_f$ flavors to one with $N_f - 1$ flavors. 
Upon giving a mass to one of the quarks in the electric theory,
\eqn\b{W={s_1\over k+1}\Tr X^{k+1} + s_2\Tr XY^2 + m\tilde Q_{N_f} Q^{N_f},}
we can integrate out the massive meson and flow to a theory with $N_f-1$
flavors. The relation between the high energy scale, $\Lambda_{N_c,N_f}$, and
the low energy scale, $\Lambda_{N_c,N_f-1}$ is
\eqn\elsc{m\Lambda_{N_c, N_f}^{N_c-N_f}=
\Lambda_{N_c, N_f-1}^{N_c-N_f+1}.}
When we give a mass to $\tilde Q_{N_f} Q^{N_f}$, we must also add the 
corresponding singlet field on the magnetic side:
\eqn\f{W=s_1\Tr \bar X^{k+1} + s_2\Tr \bar X \bar Y^2 +
{s_1 s_2\over \mu^4} \sum_{l=1}^{k} \sum_{j=1}^{3} M_{lj}\tilde q
 \bar X^{k-l} \bar Y^{3-j} q + m(M_{1,1})_{N_f}^{N_f}.}
Upon integrating out the singlet field $M$, we find that one of the mesons
on the dual side has acquired a non-zero vacuum expectation value.
\eqn\expval{\tilde q^{N_f}\bar X^{k-1}\bar Y^2q_{N_f}=-{m\mu^4\over s_1 s_2}}
For $k=2$, a solution that satisfies the F-terms, D-terms, and equation 
\expval, while giving no other meson a mass is 
$$\bar X=\left(\matrix{0&\sqrt2&0&0&0&0&0&.\cr
 		  	  0&0&0&0&1&0&0&.\cr
			  0&0&0&-1&0&0&0&.\cr
			  0&0&0&0&0&0&0&.\cr
			  0&0&0&0&0&\sqrt2&0&.\cr
			  0&0&0&0&0&0&0&.\cr
			  0&0&0&0&0&0&0&.\cr
			  .&.&.&.&.&.&.&.\cr}\right)$$
$$\bar Y=\left(\matrix{0&0&\sqrt2&0&0&0&0&.\cr
			  0&0&0&1&0&0&0&.\cr
			  0&0&0&0&1&0&0&.\cr
			  0&0&0&0&0&\sqrt2&0&.\cr
			  0&0&0&0&0&0&0&.\cr
			  0&0&0&0&0&0&0&.\cr
			  0&0&0&0&0&0&0&.\cr
			  .&.&.&.&.&.&.&.\cr}\right)$$
with $\tilde q^{N_f}_{\alpha} = 2\delta_{\alpha,1}$
and $q^{\alpha}_{N_f} = 2\delta^{\alpha,6}$.
For arbitrary $k$, the solution will take the form $\bar X = 
\bar X^{\alpha}_{\beta - 1}$
with $\bar X^k_{k+1}=0,\bar X^{2k}_{2k+1}=0$, and all the other
elements zero except for $\bar X^{k}_{2k+1}$; $\bar Y = \bar
Y^{\alpha}_{\beta - k}$ and $\tilde q^{N_f}_{\alpha} =
\delta_{\alpha,1}$ and $q^{\alpha}_{N_f} = \delta^{\alpha,3k}$ with
all other elements zero. Given this ansatz it is quite easy to insert
it into \expval, \eom, and the D-flat equations to determine the
values of non-zero matrix elements exactly up to gauge
transformations.  The exact solution for $k=3$ is given in the
appendix.  The vevs for $\bar X$ and $\bar Y$, Higgses the gauge group
from $SU(\bar N_c)$ down to $SU(\bar N_c- 3k)$.  The $q_{N_f}$ and
$\tilde q_{N_f}$ fields that received vevs will also be eaten.  The
adjoint fields $\bar X$ and $\bar Y$ will break apart into the smaller
adjoint fields of the $SU(\overline N_c - 3k)$ gauge group plus $6k$
fundamentals, $3k-1$ of which will be eaten by the Higgs mechanism and
$3k+1$ of which will receive a mass.  Thus, the magnetic theory will
flow at low energies to a theory that is dual to the low energy
electric theory.

In order to simplify the scale matching calculations, it is convenient
to make the assumption that the vacuum expectation value for $X$ and
$Y$ are the same. This calculation can also be done when $X$ and $Y$
do not have the same vev.  {}From equation \eom, this assumption
implies a relation between $s_1$ and $s_2$:
\eqn\ss{s_1 = s_2\left({m\mu^4\over s_1s_2}\right)^{2-k\over k+3}.}

The scale matching works in three stages. At the first stage, the
gauge vector bosons acquire a mass $({m\mu^4\over s_1 s_2})^{1\over
k+3}$. Thus, the scale matching for this stage is
\eqn\stgone{\bar \Lambda_{\bar N_c, N_f}^{\bar N_c-N_f}
=({m\mu^4\over s_1 s_2})^{6k\over k+3}
\bar \Lambda_{\bar N_c - 3k, N_f+3k}^{\bar N_c-6k-N_f}.}
In the second stage, the $3k+1$ fundamentals coming from the decomposition of the 
adjoint matter fields receive a mass from the superpotential terms
\eqn\massX{W_{\rm mag}={s_1\over k+1}\Tr \bar X^{k+1}\simeq
s_1\langle \bar X\rangle^{k-1} \bar X^2=s_1\left({m\mu^4\over s_1 s_2}
\right)^{k-1\over k+3}\bar X^2 = s_2\left({m\mu^4\over s_1 s_2}
\right)^{1\over k+3}\bar X^2,}
\eqn\massY{W_{\rm mag}=s_2\Tr \bar X\bar Y^2\simeq
s_2\langle \bar X\rangle \bar Y^2=s_2\left({m\mu^4\over s_1 s_2}
\right)^{1\over k+3}\bar Y^2,}
and
\eqn\massY{W_{\rm mag}=s_2\Tr \bar X\bar Y^2\simeq
s_2\langle \bar Y\rangle \bar X\bar Y=s_2\left({m\mu^4\over s_1 s_2}
\right)^{1\over k+3}\bar X\bar Y,}
where we have used \ss\ in \massX.
Flowing down in energy, the scale matching relation is
\eqn\stgtwo{\bar \Lambda_{\bar N_c-6k, N_f}^{\bar N_c-6k-N_f}
=s_2^{-3k-1}\left( {m\mu^4\over s_1 s_2}\right) ^{-3k-1\over k+3}
\bar \Lambda_{\bar N_c - 3k, N_f-1}^{\bar N_c-3k-N_f+1}}
Combining equation \stgone\ and \stgtwo\ we find 
\eqn\stgthree{\bar \Lambda_{\bar N_c, N_f}^{\bar N_c-N_f}
=s_2^{-3k-1}\left( {m\mu^4\over s_1 s_2}\right) ^{3k-1\over k+3}
\bar \Lambda_{\bar N_c - 3k, N_f-1}^{\bar N_c-3k-N_f+1}.}
Rewriting this as 
\eqn\stgthreeRW{\bar \Lambda_{\bar N_c, N_f}^{\bar N_c-N_f}
=s_2^{-3k-1}\left( {m\mu^4\over s_1 s_2}\right) ^{2k-4\over k+3}
({m\mu^4\over s_1 s_2})
\bar \Lambda_{\bar N_c - 3k, N_f-1}^{\bar N_c-3k-N_f+1},}
we see that we can use \ss\ to replace 
$({m\mu^4\over s_1 s_2})^{2k-4\over k+3}$ with $\left( {s_2\over s_1}\right)
 ^2$.
Thus, we find that the scale matching relation of the low energy theory is
\eqn\mgsc{\bar \Lambda_{\bar N_c, N_f}^{\bar N_c-N_f}
=s_1^{-3}s_2^{-3k}m\mu^4
\bar \Lambda_{\bar N_c - 3k, N_f-1}^{\bar N_c-3k-N_f+1}
.}
Now, by using the scale matching relation between the electric and the 
magnetic 
theory for $N_f$ flavors \scmat, the relation between the electric theory 
scale 
with $N_f$ flavors and $N_f - 1$ flavors 
\elsc, and the relation between the magnetic theory scale with $N_f$ flavors
and $N_f - 1$ flavors\mgsc, we can derive a relation between the electric 
theory
scale and the magnetic theory scale with $N_f-1$ flavors. It is
\eqn\scmatlow{\Lambda^{N_c-N_f+1}\bar \Lambda^{\bar N_c-N_f+1}
=Cs_1^{-3N_f+3}s_2^{-3k(N_f-1)}\mu^{4N_f-4}} where $C$ is some
constant.  By comparing with \scmat\ , we see easily the scale
matching relation has been preserved under the renormalization group
flow.

The duality map between baryons \barmap\ should more properly be
accompanied by a scale matching relation that can be determined from
the global symmetries and the flow described above.
\eqn\bmap{\eqalign{&B^{i_1\cdots i_{n_{1,1}},j_1\cdots j_{n_{2,1}},\cdots,
z_1\cdots z_{n_{k,3}}}= 
P\left(\prod_{i=1}^{3}\prod_{j=1}^{k}{1\over \bar n_{i,j}!}\right)
s_1^{N_c+3N_fk\over 2}s_2^{{3k\over 2}(N_c+N_f)-N_c}
\Lambda^{{3k\over 2}(N_c-N_f)}\mu^{-2\overline N_c}\cr
&\epsilon^{i_1\cdots i_{n_{1,1}},\bar z_1\cdots
\bar z_{\bar n_{k,3}}}\epsilon^{j_1\cdots j_{n_{2,1}},
\bar y_1\cdots\bar y_{\bar n_{k-1,3}}}
... \epsilon^{z_1\cdots z_{n_{k,3}},\bar i_1\cdots
\bar i_{\bar n_{1,1}}}
\bar B_{\bar i_1\cdots \bar i_{\bar n_{1,1}},\cdots,
\bar y_1\cdots \bar y_{\bar n_{k-1,3}},
\bar z_1\cdots \bar z_{\bar n_{k,3}}}\cr}}
$P$ is some phase.
When a flavor on the electric side is given a mass, those baryons that 
contained that quark become heavy and are not present in the low energy 
theory. On the dual side the gauge group breaks, and to preserve the 
baryon mapping, we must saturate, the first $3k$ gauge indices,
with the expectation values of the fields. Thus the mapping between the 
high energy and low energy baryons on the dual side becomes
\eqn\barflow{ \eqalign{&\epsilon^{i_1\cdots i_{n_{1,1}},\bar z_1\cdots
\bar z_{\bar n_{k,3}}}\epsilon^{j_1\cdots j_{n_{2,1}},
\bar y_1\cdots\bar y_{\bar n_{k-1,3}}}
... \epsilon^{z_1\cdots z_{n_{k,3}},\bar i_1\cdots \bar i_{\bar n_{1,1}}}
\bar B^{\bar N_c, N_f}
_{\bar i_1\cdots \bar i_{\bar n_{1,1}},\cdots, \bar y_1\cdots \bar y_{\bar
n_{1,1}},
\bar z_1\cdots \bar z_{\bar n_{k,2}}}\to  \cr    
&\bar n_{1,1}\bar n_{2,1}\cdots \bar n_{k,3}\left({m\mu^4\over s_1
s_2}\right)^{3k\over 2} 
\epsilon^{i_1\cdots i_{n_{1,1}},\bar z_1\cdots
\bar z_{\bar n_{k,3}-1}}\cdots
\epsilon^{z_1\cdots z_{n_{k,3}},\bar i_1\cdots \bar i_{\bar n_{1,1}-1}}
\bar B^{\bar N_c-6, N_f-1}
_{\bar i_1\cdots \bar i_{\bar n_{1,1}-1},\cdots,
\bar z_1\cdots \bar z_{\bar n_{k,3}-1}}\cr}}
Using equation \elsc\ we see that these 
relations are just right to preserve the duality mapping between baryonic 
operators \bmap\ under the renormalization group flow from the theory with 
$N_f$ flavor to the theory with $N_f - 1$ flavors.

\subsec{Flat directions of adjoint fields for odd values of k}

It is convenient to consider the theories with even and odd values of $k$
separately. We consider the $k$ odd case first.
\eqn\supm{W={s_1\over k+1}\Tr X^{k+1} + s_2\Tr XY^2
- \lambda_1 \Tr X - \lambda_2 \Tr Y.}
The equations of motion 
\eqn\eom{\eqalign{s_1 X^k + s_2 Y^2 - \lambda_1 = 0\cr
XY+YX - \lambda_2 = 0. \cr}}
An $SU(2n+km)$ gauge theory will have a flat direction along which X gets a 
vacuum expectation value
$$X=a\left(\matrix{0_n&0&0&0&0&0&.&0\cr
                  0&0_n&0&0&0&0&.&0\cr
                  0&0&1_m&0&0&0&.&0\cr
                  0&0&0&\omega_m &0&0&.&0\cr
                  0&0&0&0&\omega_m^2&0&.&0\cr
                  0&0&0&0&0&\omega_m^3&.&0\cr
		  .&.&.&.&.&.&.&.\cr
		  0&0&0&0&0&0&.&\omega_m^{k-1}\cr}\right)$$
$$Y=b\left(\matrix{1_n&0&0&0&0&0&.&0\cr
                  0&-1_n&0&0&0&0&.&0\cr
                  0&0&0_m&0&0&0&.&0\cr
                  0&0&0&0_m&0&0&.&0\cr
                  0&0&0&0&0_m&0&.&0\cr
                  0&0&0&0&0&0_m&.&0\cr
                  .&.&.&.&.&.&.&.\cr
                  0&0&0&0&0&0&.&0_m\cr}\right),$$
where $\omega=\exp{2\pi i\over k}$ and 
$a=\root k \of {\lambda_1\over s_1}$ and $b=\root 2 \of {\lambda_1\over s_2}$.
The subscripts on the entries indicate that they
are matrices.
One could think of the $Y$ field's 
vacuum
expectation value breaking the theory into three, and then the $X$ field 
vev breaking one of those three into $k$ parts. We end up with 
$SU(n)\times SU(n)\times SU(m)^k\times U(1)^{k+1}$.
The fields $X$ and $Y$ each decompose into $k+2$ adjoints and fields in the 
$(n,n)$, $(n,m)$, $(m,n)$, and $(m,m)$ representations. We will call the 
fields, coming from $X$ in the 
$(n,n)$ representation, $F$. It can be shown $F$ does not receive a mass 
from the superpotential and that there is the term 
\eqn\intW{W_L = {s_1\over k+1}\Tr (F\tilde F)^{k+1\over 2}}
All the other matter fields (except the $Q$s) will receive a mass upon
decomposition of the superpotential.
Thus along this flat direction the theory flows to a duality discussed in 
a paper by 
Intriligator, Leigh, and Strassler \ilst\ in 
the $SU(n)\times SU(n)$ part 
and Seiberg's SQCD \nati\ in the 
$SU(m)^k$ part. On the dual side, the dual gauge group was 
$SU(3kN_f - N_c)$. It breaks to 
$SU(kN_f-n)\times SU(kN_f-n)\times SU(N_f-m)^k\times U(1)^{k+1}$ which is just 
right for preserving the duality. 

It can be shown that the scale matching relation \scmat\ and 
the scale matching relation for SQCD are consistent with this flat direction.

It was explained in \ilst\ how to deform
\intW\ by lower order operators and flow to SQCD. One can think of deformations
 of \intW\ in the high energy theory as adding even powered operators $\Tr X^r$
 to the superpotential \supm .

\subsec{Deformations of superpotential by adjoint fields for $k$ even.}

We now move on and look at the case where $k$ is even. {}From
section 2, we remember that this was the case in which the classical 
superpotential did not truncate the chiral ring sufficiently for the proposed 
duality to be valid. Assuming duality, additional constraints must be present.
To implement these constraints we add to our superpotential some 
Lagrange multipliers, setting in particular the operators $\Tr Y^{j}$ to zero 
where $j>2$. The superpotential looks like
\eqn\etaW{W = s_1 \Tr X^{k+1} + \Tr XY^2 - \lambda_1 \Tr X - \lambda_2 \Tr Y + 
	{\eta_3\over 3} \Tr Y^3 + {\eta_4\over 4} \Tr Y^4 + \cdots}
where the $\eta$s are the Lagrange multipliers.
Let's now deform the superpotential by $X$ operators of odd powers only
(that are in this modified chiral ring).
The superpotential becomes
\eqn\defW{W = \sum_{r=1}^{k\over 2} {g_r\over 2r+1} \Tr X^{2r+1} + \Tr XY^2 
 - \lambda_1 \Tr X - \lambda_2 \Tr Y +  {\eta_3\over 3} \Tr Y^3 
 + {\eta_4\over 4} \Tr Y^4 + \cdots}
The equations of motion from \defW\ are
\eqn\eomeven{\eqalign{\sum_{r=1}^{k\over 2} g_r X^{2r} + s_2 Y^2 - \lambda_1 = 0\cr
XY+YX -\lambda_2 + \eta_3 Y^2 + \eta_4 Y^3 + \cdots = 0. \cr}}
Because we deformed the superpotential by lower order terms of odd powers only,
 we see that eigenvalues for $X$, with $Y=0$, will come in ${k\over 2}$ plus 
and minus pairs. 
We can now consider tuning the $g_r$
coefficients in the superpotential \defW\ such that there is only one pair of
plus and minus roots
$$X=\left(\matrix{a&0\cr
                  0&-a\cr}\right),$$
where $a$ depends on the $g_r$s.
The vacuum expectation value of $X$ breaks the gauge group into two parts
$SU({N_c\over 2})\times SU({N_c\over 2})\times U(1)$.
The dual gauge group, because traces of products of $X$ and $Y$ map to the 
same traces of products of $\bar X$ and $\bar Y$, breaks to
$SU({3kN_f-N_c\over 2})\times SU({3kN_f-N_c\over 2})\times U(1)$. The
field $X$ will decompose into an adjoint, we will call, $X_1$ charged under 
the first gauge
group, an adjoint $X_2$ charged under the second gauge group, and some
fields that are eaten by the Higgs mechanism. The field
$Y$, however, decomposes into adjoints $Y_1$ and $Y_2$ and fields $F$ and
$\tilde F$
that are not eaten and are in the $({N_c\over 2},{N_c\over 2})$ representation 
of the product gauge group.
Under this decomposition, the superpotential becomes
\eqn\supbrk{\eqalign{W =  & s_1(2a)^{{k\over 2}} \Tr X_1^{{k\over 2}+1}
+ s_1(-2a)^{{k\over 2}} \Tr X_2^{{k\over 2}+1} 
 + s_2a\Tr Y_1^2 - s_2a\Tr Y_2^2 \cr
+ & s_2\Tr X_1Y_1^2 + s_2\Tr X_2Y_2^2 + s_2\Tr X_1F\tilde F + s_2\Tr X_2F\tilde F 
 + \eta_3 \Tr Y_1F\tilde F \cr
+ & \eta_3 \Tr Y_2\tilde FF + {\eta_3 \Tr Y_1^3\over 3} 
+ {\eta_3 \Tr Y_2^3\over 3} 
 + \eta_4 \Tr (F\tilde F)^2  + {\eta_4 \Tr Y_1^4\over 4} 
+ {\eta_4 \Tr Y_2^4\over 4} \cr
 + & \eta_4 \Tr Y_1^2F\tilde F + \eta_4 \Tr Y_2^2\tilde FF + \cdots}}
We see that the fields $F$ do not get a mass.
We can integrate out the massive fields $Y_1$ and $Y_2$, and we are
left with
\eqn\supbrk{\eqalign{W = & s_1(2a)^{{k\over 2}} \Tr X_1^{{k\over 2}+1}
+ s_1(-2a)^{{k\over 2}} \Tr X_2^{{k\over 2}+1}\cr
+ & s_2\Tr X_1F\tilde F + s_2\Tr X_2F\tilde F + \eta_4 \Tr (F\tilde F)^2.\cr}}
Thus, by deforming the superpotential and flowing to a lower energy theory, we
have been led to a prediction for a new duality. In fact, it can be shown
that the anomaly matching conditions still holds for the theory with
superpotential \supbrk. As in the high energy theory, the chiral ring is 
not truncated by the classical equations of motion in the way necessary for 
the duality. The necessary additional constraints, in the form of 
the Lagrange multipliers such as $\eta_4$, were implemented in the 
high energy theory. The difference, noted earlier, between the $k$ 
even and $k$ odd cases has followed us in the renormalization group flow
as might be expected.

We can now go back to superpotential \defW\ and ask what happens more
generally when there are ${k\over 2}$ distinct plus and minus pairs of roots.
The theory will break apart into ${k\over 2}$ decoupled copies of the above 
described
duality. The superpotential in one of these ${k\over 2}$ vacua is 
\eqn\supbrkone{\eqalign{W = & s_1(2a)\prod_{i=1}^{{k\over 2}-1}(a-b_i)(a+b_i) \Tr X_1^2
+ s_1(-2a)\prod_{i=1}^{{k\over 2}-1}(a-b_i)(a+b_i) \Tr X_2^2 \cr
+ & s_2\Tr X_1F\tilde F + s_2\Tr X_2F\tilde F + \eta_4 \Tr (F\tilde F)^2.\cr}}
Where $a$ and $b_i$ depend on the $g_r$s in \defW. 
The electric gauge group breaks from $SU(N_c)$ down to 
$\prod_{i=1}^{{k\over 2}} SU(n_i)\times SU(n_i)$ while the magnetic gauge group break from
$SU(3kN_f - N_c)$ down to 
$\prod_{i=1}^{{k\over 2}} SU(3N_f - n_i)\times SU(3N_f - n_i).$
Now, we can integrate out 
the massive fields $X_1$ and $X_2$, and we are left with
\eqn\etaL{W_L = \eta_4 \Tr (F\tilde F)^2} which is just a superpotential
of the Intriligator-Leigh-Strassler duality mentioned in the previous 
subsection. We can now add a mass term to the superpotential, 
\eqn\etamass{W_L = \eta_4 \Tr (F\tilde F)^2 + m\Tr F\tilde F,}
which breaks the theory down to $SU(p_0)\times U(p)\times SU(p_0)$ where 
there are $N_f$ flavors charged under
the $SU(p_0)$s and $2N_f$ flavors charged under the diagonal subgroup $U(p)$
The dual theory breaks to $SU(N_f - p_0)\times U(2N_f - p)\times SU(N_f - p_0)$
\ilst. There is an adjoint field charged under the diagonal subgroup
$U(p)$ which must receive a mass if the duality is to be that of 
SQCD \nati. We see that the mass comes from the Lagrange multiplier
term. Thus, the duality of the $k$ evens is consistent
if we impose additional constraints, not implied by the classical equations 
of motion.  

\newsec{A New Duality}

This section can be read independently of the other sections.
The deformations of the even $k$ superpotential in the previous section led 
us to a prediction for a new duality. 
This duality can be generalized to one in which the electric theory has an 
$SU(N_c)\times SU(N_c')$ gauge group with $N_f$
fundamentals $Q$ and $N_f$ anti--fundamentals $\tilde Q$
charged under only $SU(N_c)$, $N_f'$ fundamentals $Q'$ and $N_f'$ 
anti--fundamentals $\tilde Q'$ charged only under
$SU(N_c')$, a field $F$ which is in an $(N_c,N_c')$ representation, 
a field $\tilde F$ which is in an $(\bar N_c,\bar N_c')$ representation, 
a field $X_1$ which is in the adjoint representation of $SU(N_c)$, and a 
field $X_2$ which is in the adjoint representation of $SU(N_c')$.

\thicksize=1pt
\vskip12pt
\begintable
\tstrut  | $\Ql;\Qlt$ | $\Qr;\Qrt$ | $F;\tilde F$ | $X_1$ | $X_2$ \crthick
$SU(\ncl)$ | ${\bf\ncl;\overline\ncl}$ | ${\bf 1;1}$ |
${\bf\ncl;\overline\ncl}$ | adj | 1  \cr
$SU(\ncr)$ | ${\bf 1;1}$ | ${\bf\ncr;\overline\ncr}$ |
${\bf\ncr;\overline\ncr}$ | 1 | adj \cr
$SU(\nfl)_L$ | ${\bf\nfl;1}$ | ${\bf 1;1}$ | ${\bf 1;1}$ | 1 | 1 \cr
$SU(\nfr)_L$ | ${\bf 1;1}$ | ${\bf\nfr;1}$ | ${\bf 1;1}$ | 1 | 1\cr
$SU(\nfl)_R$ | ${\bf 1;\nfl}$ | ${\bf 1;1}$ | ${\bf 1;1}$ | 1 | 1\cr
$SU(\nfr)_R$ | ${\bf 1;1}$ | ${\bf 1;\nfr}$ | ${\bf 1;1}$ | 1 | 1\cr
$U(1)_B$ | $\pm {1\over\ncl}$ | $\mp {1\over\ncr}$ |
$\mp \left({1\over\ncl}+{1\over\ncr}\right)$ | 0 | 0 \cr
$U(1)_{B'}$ | $\pm {1\over\ncl}$ | $\pm {1\over\ncr}$ |
$\mp \left({1\over\ncl}-{1\over\ncr}\right)$ | 0 | 0 \cr
$U(1)_F$ | 0 | 0 | $\pm 1$ | 0 | 0 \cr
$U(1)_R$ | $1+{(\ncr-2\ncl)\over\nfl (k+1)}$ |
$1+{(\ncl-2\ncr)\over\nfr (k+1)}$ |
${k\over k+1}$ | ${2\over k+1}$ | ${2\over k+1}$
\endtable
\noindent

The superpotential is 
\eqn\supND{\eqalign{W = & {s_1\over k+1}\Tr X_1^{k+1}
+ {s_1(-1)^{k+1}\over k+1}\Tr X_2^{k+1}\cr
+ & s_2\Tr X_1F\tilde F + s_2\Tr X_2F\tilde F + \lambda_1\Tr  X_1 
+ \lambda_2 \Tr X_2 \cr}}
where the $\lambda$s are there to enforce the tracelessness of the adjoints.
The equations of motion are now
\eqn\eomND{\eqalign{& s_1 X_1^k  + s_2F\tilde F + \lambda_1 = 0\cr
 & s_1(-1)^{k+1} X_2^k + s_2 F\tilde F + \lambda_2 = 0 \cr
 & X_1F + X_2F = 0 \cr
 & X_1\tilde F + X_2\tilde F = 0.\cr}}
The equations of motion truncate the chiral ring in a manner similar to that 
discussed in section 2 for the $k$ odd two adjoint case. Dropping the 
$\lambda$s and multiplying the 
first equation of motion \eomND\ by $\tilde F$, we find
\eqn\ringND{s_1 X_1^k \tilde F  = - s_2\tilde F F\tilde F.}
The second equation effectively turns $X_1$s into $X_2$s leaving a minus sign
each time. We then have
\eqn\ringsND{s_1(-1)^k X_2^k \tilde F  = - s_2\tilde F F\tilde F;} 
comparing this with the last equation of motion in \eomND\ we see that 
$\tilde FF\tilde F = 0$. By a similar calculation one can show that 
$F\tilde FF = 0$. 

Owing to the superpotential, the mesons in the theory are  
$M_{l,1} = QX_1^{l-1}\tilde Q$, $P_{l,1} = Q'X_2^{l-1}\tilde Q'$, 
$M_{l,2} = QX_1^{l-1}F\tilde Q'$,
$P_{l,2} = Q'X_2^{l-1}\tilde F\tilde Q$,
$M_{l,3} = QF\tilde FX_1^{l-1}\tilde Q$, 
$P_{l,3} = Q'\tilde FFX_2^{l-1}\tilde Q'$,
where $j=1\cdots k$

We can also form baryons by introducing dressed quarks
\eqn\j{\eqalign{Q_{(l,1)}= & X_1^{l-1}Q \cr
		Q_{(l,2)}= & X_1^{l-1}\tilde FQ'\cr
		Q_{(l,3)}= & X_1^{l-1}\tilde FFQ\cr
		Q'_{(l,1)}= & X_2^{l-1}Q'\cr
		Q'_{(l,2)}= & X_2^{l-1}FQ\cr
		Q'_{(l,3)}= & X_2^{l-1}F\tilde FQ';\;\;\;l=1,\cdots k.}}
and then contracting the gauge indices on an $SU(N_c)$ 
epsilon tensor,
\eqn\baryon{B^{(n_{1,1},n_{2,1},\cdots,n_{k,3})}=Q_{(1,1)}^{n_{1,1}}\cdots
Q_{(k,3)}^{n_{k,3}}
;\;\;\sum_{l=1}^{k} \sum_{j=1}^{3} n_{l,j}=N_c,}
or on an $SU(N_c')$
epsilon tensor to form
\eqn\baryon{{B'}^{(n'_{1,1},n'_{2,1},\cdots,n'_{k,3})}={Q'}_{(1,1)}^{n'_{1,1}}
\cdots
{Q'}_{(k,3)}^{n'_{k,3}}
;\;\;\sum_{l=1}^{k} \sum_{j=1}^{3} n'_{l,j}=N_c'.}
The total number of baryons is
\eqn\totnum{\eqalign{\sum_{\{n_{l,j}\}}{N_f\choose n_{1,1}}\cdots{N_f'\choose n_{k,2}}
\cdots{N_f\choose n_{k,3}} + & \sum_{\{n'_{l,j}\}}{N_f'\choose n'_{1,1}}
\cdots{N_f\choose n'_{k,2}}\cdots{N_f'\choose n'_{k,3}} \cr
= & {2kN_f + kN_f'\choose N_c} + {2kN_f' + kN_f\choose N_c'.}}}
There are also anti-baryons and traces of the form $\Tr X_1^r$ and 
$\Tr X_2^r$.

\subsec{duality}
The dual theory has a gauge group 
$SU(2kN_f' + kN_f - N_c')\times SU(2kN_f + kN_f' - N_c)$
with matter content :

\thicksize=1pt
\vskip12pt
\begintable
\tstrut  | $\ql;\qlt$ | $\qr;\qrt$ | $\bar F;\tilde {\bar F}$ | $\bar X_1$ | $\bar X_2$ \crthick
$SU(\ncld)$ | ${\bf\ncld;\overline\ncld}$ | ${\bf 1;1}$ |
${\bf\ncld;\overline\ncld}$ | adj | 1 \cr
$SU(\ncrd)$ | ${\bf 1; 1}$ | ${\bf\ncrd;\overline\ncrd}$ |
${\bf \ncrd;\overline\ncrd}$ | 1 | adj \cr
$SU(\nfl)_L$ | ${\bf 1;1}$ | ${\bf \overline\nfl; 1}$ | ${\bf
1;1}$ | 1 | 1 \cr
$SU(\nfl)_R$ | ${\bf 1;1}$ | ${\bf 1; \overline\nfl}$ | ${\bf
1;1}$ | 1 | 1 \cr
$SU(\nfr)_L$ | ${\bf \overline\nfr;1}$ | ${\bf 1; 1}$ | ${\bf
1;1}$ | 1 | 1 \cr
$SU(\nfr)_R$ | ${\bf 1;\overline\nfr}$ | ${\bf 1; 1}$ | ${\bf
1;1}$ | 1 | 1 \cr
$U(1)_B$ | $\mp {1\over\ncld}$ | $\pm {1\over\ncrd}$ |
 $\pm \left({1\over\ncld}+{1\over\ncrd}\right)$ | 0 | 0 \cr
$U(1)_{B'}$ | $\pm {1\over\ncld}$ | $\pm {1\over\ncrd}$ |
$\mp\left({1\over\ncld}-{1\over\ncrd}\right)$ | 0 | 0 \cr
$U(1)_F$ | $\mp {\nfl\over\ncld}$ | $\pm {\nfr\over\ncrd}$
  | $\mp\left(1-{\nfl\over\ncld}-{\nfr\over\ncrd}\right)$ | 0 | 0 \cr
$U(1)_R$ | $1+{(\ncrd-2\ncld)\over\nfr (k+1)}$ |
$1+{(\ncld-2\ncrd)\over\nfl (k+1)}$ | ${k\over k+1}$ | ${2\over k+1}$ | 
${2\over k+1} $
\endtable
\noindent

and a dual superpotential of the form
\eqn\supND{\eqalign{W= & {s_1\over k+1}\Tr \bar X_1^{k+1}
+  {s_1(-1)^{k+1}\over k+1}\Tr \bar X_2^{k+1} + s_2\Tr \bar X_1\bar F\tilde F 
+ s_2\Tr \bar X_2\bar F\tilde {\bar F} + \lambda_1 \Tr \bar X_1 
+ \lambda_2 \Tr \bar X_2 \cr
+ & {s_1s_2\over \mu^4}\sum_{l=1}^{k} 
    M_{l,1}\tilde q' \bar X_2^{k-l} \bar F \tilde {\bar F} q'
+   P_{l,1}\tilde q  \bar X_1^{k-l} \tilde {\bar F} \bar F q 
+   P_{l,2}\tilde q' \bar X_2^{k-l} \bar F q \cr
+ & M_{l,2}\tilde q  \bar X_1^{k-l} \tilde {\bar F} q' 
+   M_{l,3}\tilde q' \bar X_2^{k-l} q'
+   P_{l,3}\tilde q  \bar X_1^{k-l} q\cr}}

The mesons are mapped to the singlets where as the
baryon-like objects are mapped to other baryon-like objects in the dual theory:
The mapping is
\eqn\barmapF{\eqalign{B_{\rm el}^{(n_{1,1},n_{2,1},\cdots,n_{k,3})}
\leftrightarrow
{B'}_{\rm mag}^{(m'_{1,1},m'_{2,1},\cdots,m'_{k,3})};\;\;
m'_{l,1}=N_f- n_{k+1-l,3};\cr
m'_{l,2}=N_f'-n_{k+1-l,2};\cr
m'_{l,3}=N_f- n_{k+1-l,1};\cr
l=1,2,\cdots, k\cr }}

\eqn\barmapFp{\eqalign{{B'}_{\rm el}^{(n'_{1,1},n'_{2,1},\cdots,n'_{k,3})}
\leftrightarrow
B_{\rm mag}^{(m_{1,1},m_{2,1},\cdots,m_{k,3})};\;\;
m_{l,1}=N_f'- n'_{k+1-l,3};\cr
m_{l,2}=N_f - n'_{k+1-l,2};\cr
m_{l,3}=N_f'- n'_{k+1-l,1};\cr
l=1,2,\cdots, k\cr }}

The fact that this mapping is consistent with all global symmetries is another
non-trivial test of the proposed duality. The t'Hooft anomaly matching conditions are satisfied for this duality.

We saw that the chiral ring was truncated in the way necessary for
the duality for all values of
$k$ provided the factor $(-1)^{k+1}$ is in front of $\Tr X_2^{k+1}$.
It would be surprising if the duality depended critically on the
numerical coefficients in the superpotential. We see again that the
duality seems to suggest that the chiral ring is truncated by some other
mechanism.

\newsec{Conclusions.}

One should make a distinction between the two cases: odd and even
$k$. For odd $k$ the two adjoint theory is has essentially the same
features as the one adjoint theory. Classically, there were many flat
directions that could be labeled by gauge invariant operators. A
superpotential was added that lifted many of them and made the theory
tractable.  The classical equations of motion of the superpotential
told us exactly which gauge invariants to keep and which ones we
should set equal to zero.  In the even $k$ case, the classical
equations of motion coming from the superpotential do not tell us
which invariants to keep. We have seen that the duality holds if many
of them are set to zero. Perhaps some of the gauge invariant mesons on
the electric side could be mapped to some of the gauge invariant
mesons on the magnetic side. In this way they would never appear as
singlets in the dual theory. Such a mapping has not been found.  It is
possible that non-perturbative effects could truncate the chiral ring
in a manner similar to that discussed in
\kss. It would be interesting to see how this would work in the case considered
here. One might speculate that a quantum truncation of the chiral ring might occur in the conformally invariant non-Abelian Coulomb phase where both 
theories are strongly coupled. The non-Abelian Coulomb phase is also the 
phase in which analogies with two dimensional conformal field theory 
would be most relevant. 

We saw that it was the $k$ even theories that led us to the new
duality.  However, in section 5, this duality was shown to be more
general than than was evident when embedded in the $k$ even theories,
and in some instances the chiral ring can be shown to truncate
classically.

Hopefully, this example will help us understand more about duality in 
supersymmetric gauge theories.

\centerline{{\bf Acknowledgments}}

I would like to thank E. Witten, D. Kutasov, W. Taylor and especially 
K. Intriligator for helpful discussions. This work was supported by NSF grant
PHY90-21984.

\newsec{Appendix.}
Here we present a D-flat solution to \expval\ for the $k=3$ theory. It is
$$X=\left(\matrix{0&\sqrt17&0&0&0&0&0&0&0&.\cr
		  0&0&-4&0&0&0&0&0&0&.\cr
		  0&0&0&0&0&0&{3\sqrt7\over2}&0&0&.\cr
		  0&0&0&0&-2&0&0&0&0&.\cr
		  0&0&0&0&0&-2&0&0&0&.\cr
		  0&0&0&0&0&0&0&0&0&.\cr
		  0&0&0&0&0&0&0&4&0&.\cr
		  0&0&0&0&0&0&0&0&\sqrt17&.\cr
		  0&0&0&0&0&0&0&0&0&.\cr
		  .&.&.&.&.&.&.&.&.&.\cr}\right)$$
$$Y=\left(\matrix{0&0&0&{\sqrt17\over 2}&0&0&0&0&0\cr
                  0&0&0&0&1&0&0&0&0&.\cr
                  0&0&0&0&0&{-1\over 2}&0&0&0&.\cr
                  0&0&0&0&0&0&{-1\over 2}&0&0&.\cr
                  0&0&0&0&0&0&0&-1&0&.\cr
                  0&0&0&0&0&0&0&0&-{\sqrt17\over 2}&.\cr
                  0&0&0&0&0&0&0&0&0&.\cr
                  0&0&0&0&0&0&0&0&0&.\cr
                  0&0&0&0&0&0&0&0&0&.\cr
		  .&.&.&.&.&.&.&.&.&.\cr}\right)$$
with $\tilde q^{N_f}_{\alpha} = {\sqrt 85\over 2}\delta_{\alpha,1}$
and $q^{\alpha}_{N_f} = {\sqrt 85\over 2}\delta^{\alpha,9}$.

\listrefs
\bye